# Ultrastable Metallic Glass Nanoparticles with Size-Dependent Mechanical Properties

Abhinav Parakh[1], Mehrdad T. Kiani[1], Anabelle Colmenares[2], Andrew C. Lee[1], Guoyin Shen[3], Stella Chariton[4], Vitali B. Prakapenka[4], and X. Wendy Gu[2]*

[1]Materials Science and Engineering, Stanford University, Stanford, CA 94305, USA

[2]Mechanical Engineering, Stanford University, Stanford, CA 94305, USA

[3]HPCAT, X-ray Science Division, Argonne National Laboratory, Lemont, IL 60439, USA

[4]Center for Advanced Radiations Sources, The University of Chicago, Chicago, IL 60637, USA

*corresponding author: xwgu@stanford.edu

**Abstract:** The atomistic structure of metallic glasses is closely related to properties such as strength and ductility. Here, $Ni_{1-x}B_x$ metallic glass nanoparticles of two different sizes are compressed under quasi-hydrostatic high-pressure conditions in order to understand structural changes under stress. The structural changes in the nanoparticles were tracked using in situ high-pressure X-ray diffraction (XRD). The ambient pressure pair-distribution functions generated from XRD showed that the smaller sized nanoparticles had a more compact amorphous structure with lower coordination number. XRD showed that the amorphous structure was stable up to the maximum pressures achieved. The bulk modulus of the smaller and larger sized nanoparticles was found to be 208 GPa and 178 GPa, respectively. This size-dependent high-pressure behavior was related to compositional differences between the nanoparticles. These results show that $Ni_{1-x}B_x$ metallic glass nanoparticles are highly stable under pressure, which could enable their use as inclusions in metal or ceramic matrix composites.

Metallic glasses (MGs) are metallic alloys with an amorphous structure, that typically exhibit high strength, chemical stability, corrosion resistance, and radiation resistance[1]. These properties are linked to structural parameters such as free volume, short and medium range order and the distribution of atomistic clusters. Changes in these structural parameters under stress are particularly important for understanding deformation mechanisms (e.g. plasticity through shear transformation zones)[2] in metallic glasses that are used as structural materials or protective coatings. Compression under high hydrostatic pressure in combination with in-situ X-ray characterization can be used to measure subtle changes in the atomistic structure of MGs under stress and explore the structural stability of the amorphous phase[3,4]. In addition, these experiments provide a highly accurate measurement of bulk modulus which is used as a predictor of ductility (along with shear modulus)[5]. Pressure induces crystallization of amorphous materials and give an upper limit on the stability of MGs under operational stresses. Zheng et al. observed pressure induced crystallization of $Ce_{0.8}Al_{0.2}$ MG at ~37 GPa and $Ce_{0.75}Al_{0.25}$ MG at ~25 GPa under helium pressure medium[6,7]. A reversible phase transition between amorphous and crystalline phases has also been observed in $As_2Se_3$ (at ~44 GPa) and GaSb (at ~5 GPa)[8,9]. Pressure induced changes in the short-range order in oxide and metallic glasses have also been observed. $TiO_2$ transformed from a low density amorphous phase to a high density amorphous phase with increasing pressure[10,11]. Pressure has also driven an increase in coordination number in amorphous $GeO_2$[12]. Dziegielewski et al. showed that the increasing hydrostatic pressure modified the local atomic configurations in Zr-Cu MG to form Cu-centered icosahedra as the dominant short-range order structural motif[13].

Here, we use X-ray diffraction (XRD) and high-pressure techniques on 41 nm and 164 nm $Ni_{1-x}B_x$ nanoparticles made via colloidal synthesis. These nanoparticles have previously been shown to be highly ductile under uniaxial compression compared to other nanoscale metallic glasses[14]. Under uniaxial compression, the smaller nanoparticles deformed by a slowly propagating shear band whereas the larger particles deformed homogenously through gradual shape change. The propensity for shear banding decreased with increasing particle size, indicating a brittle-to-ductile transition with increasing size which is opposite to the trend for nanoscale MGs fabricated from bulk metallic glasses[15–17]. We investigate how these properties are related to the ambient pressure atomistic structure, changes in atomistic structure under hydrostatic stress, and the cluster-based growth mechanism, which differs from other metallic glass fabrication techniques[18]. The ambient pressure XRD showed a difference in the coordination number and bonding distances for the 41 nm and 164 nm nanoparticles. A lower coordination number and smaller nearest neighbor spacing for the 41 nm nanoparticles was related to the higher boron content in the smaller nanoparticles than the 164 nm nanoparticles. The nanoparticles were pressurized to high pressures using neon quasi-hydrostatic pressure medium in a diamond anvil cell (DAC). In situ XRD was performed to track the changes in the reduced volume and pair-distribution functions with increasing pressure. XRD data was used to calculate the bulk modulus of the nanoparticles which was observed to be larger for the 41 nm than for the 164 nm nanoparticles. The high-pressure deformation showed no phase transformation up to ~55 GPa for 41 nm nanoparticles and up to ~41 GPa for 164 nm nanoparticles. This showed that the nanoparticles were extremely stable under pressure. This work provides insight into the deformation of MG nanoparticles at high pressures, which may be relevant to the use of the nanoparticles in composite materials or to form bulk nano-glasses[19,20].

**Experimental Methods**

*Synthesis*

$Ni_{1-x}B_x$ nanoparticles were synthesized using established methods[14,21]. Briefly, an aqueous solution of 1 mM nickel nitrate, 10 mM SDS, and 0.1 mM oleic acid was prepared at room temperature. Oleic acid was pre-mixed in methanol solution. $NaBH_4$ was rapidly added to the aqueous solution while stirring to form nanoparticles. The amount of $NaBH_4$ determined the size of the nanoparticles. Reducing the amount of $NaBH_4$ increased the size of particles formed. For example, to synthesize $41 \pm 3$ nm sized particles, 2.5 mg of $NaBH_4$ was added to 10 mL of total solution. The solution turned dark within seconds and the particles were separated and washed several times by centrifugation. Synthesis of larger nanoparticles resulted in a larger size distribution. Smaller particles were removed using size-selective centrifugation. All chemicals were purchased from Sigma-Aldrich. Transmission Electron Microscope (TEM) images were taken on a FEI Tecnai G2 F20 X-TWIN at 200 kV.

*Ambient pressure XRD*

Dried nanoparticles was transferred onto a 30 μm thick Kapton tape. Kapton tape background scattering was subtracted from the sample diffraction data. The 41 nm nanoparticle ambient pressure diffraction data was collected at beamline 16-BMD at the Advanced Photon Source, Argonne National Lab with 0.2952 Å as the X-ray wavelength and a Mar345 image plate as the detector. The 164 nm nanoparticle diffraction data was collected at beamline 13-IDD at GSECARS, Advanced Photon Source, Argonne National Lab with 0.2952 Å as the X-ray wavelength and Dectris Pilatus CdTe 1M was used as the detector.

*High pressure XRD*

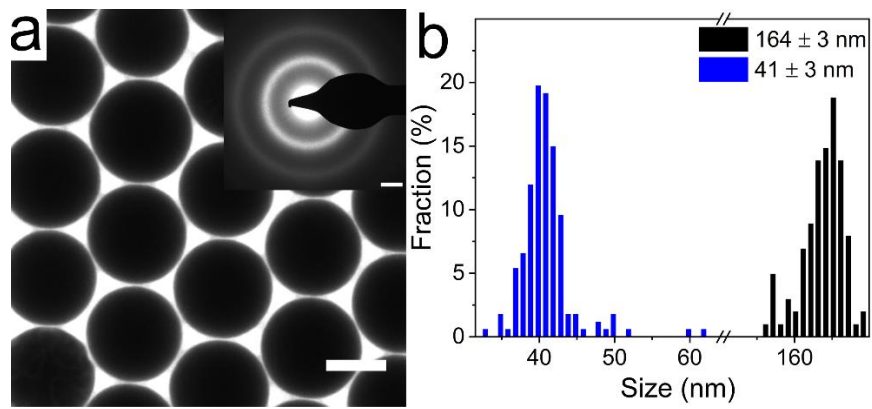

**Figure 1:** a) TEM image of 164 nm sized $Ni_{1-x}B_x$ nanoparticles. Scale bar is 100 nm. Inset showing TEM diffraction pattern for 41 nm $Ni_{1-x}B_x$ nanoparticles. Scale bar is 1 $nm^{-1}$. b) Size distribution of 41 nm and 164 nm sized $Ni_{1-x}B_x$ nanoparticles measured using TEM images.

Mao-type symmetric diamond anvil cells with 300 μm and 400 μm culets were used to conduct high pressure experiments. Re gaskets were pre-indented to 50 μm thickness and 150 μm holes were drilled using an EDM machine and used as the sample chamber. The sample was dried out on a silicon wafer and then a ~50 μm sized piece of the sample was loaded into diamond anvil cell using a needle. Ruby was used as a pressure calibrant. Neon gas was loaded as the pressure medium for quasi-hydrostatic measurements. In situ XRD experiments were conducted at Advanced Photon Source, Argonne National Lab beamline 16-BMD. An X-ray beam with a wavelength of 0.2952 Å and Mar image plate detector with an exposure time of 300 s were used for diffraction measurements. Background scattering was measured by collecting an XRD pattern on an empty location in the sample chamber.

*XRD analysis*

Radial integration was performed using the software DIOPTAS[22] after masking neon pressure medium and diamond diffraction peaks. The peak fitting to extract peak position was performed using a combination of Gaussian and Lorentzian peak profiles with a high order polynomial for the background. For calculating the pair distribution function (PDF), a careful experimental background subtraction was performed using the background diffraction collected from an empty space in the DAC sample chamber. The experimental background was scaled to match the scattering at small angles before subtracting. The secondary diffraction from the Re gasket was removed from the experimental data set by peak fitting and then removing the fitted Re peaks. This method of removing secondary diffraction peaks was developed following Hong et al. [23]. This experimental data set was then used to generate the PDFs using PDFgetX2 software. Nanoparticle composition was estimated using inductively coupled plasma optical emission spectrometry[14]. Nanoparticle composition of $Ni_{0.71}B_{0.29}$ and $Ni_{0.79}B_{0.21}$ was used for 41 nm and 164 nm nanoparticles, respectively. Radial distribution function was derived from the PDF to calculate the coordination number[24].

**Results**

MG nanoparticles were synthesized in aqueous solution by reducing Ni metal ions using $NaBH_4$ as a reducing agent. Nanoparticle size was controlled by the amount of $NaBH_4$. Reducing the amount of $NaBH_4$ limits the number of nucleation sites which enabled the growth of $Ni_{1-x}B_x$ nanoparticles to larger sizes. $NaBH_4$ concentration influences the boron concentration in the nanoparticles. In a previous study, 80-100 nm nanoparticles were found to have a lower boron content of ~21% and 20-30 nm nanoparticles were found to have a higher boron content of ~29%[14]. TEM images of the nanoparticles showed that the synthesized nanoparticles were spherical with a monomodal size distribution (see Fig 1a). The corresponding electron diffraction pattern for the nanoparticles show amorphous diffraction rings (Fig 1a inset). The size distribution was found to be 41 ± 3 nm and 164 ± 3 nm with no overlap in observed particle sizes (see Fig. 1b).

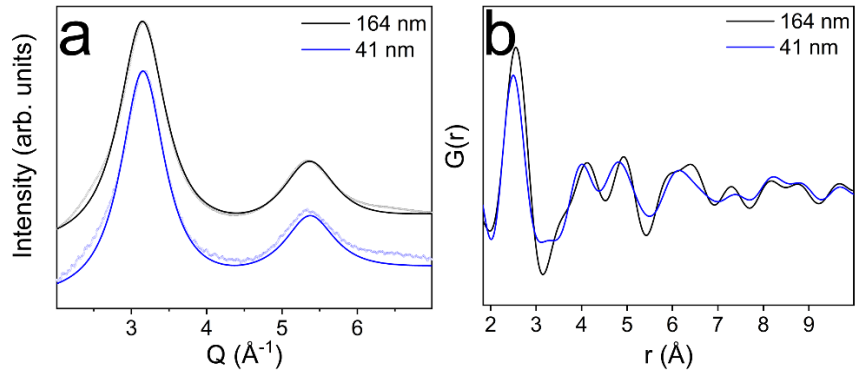

**Figure 2:** a) Experimental high pressure in situ XRD patterns and corresponding peak fitting for the amorphous peaks. b) PDF at ambient pressure for 41 nm and 164 nm $Ni_{1-x}B_x$ nanoparticles.

Fig. 2a shows the ambient pressure XRD for 41 nm and 164 nm nanoparticles. The first XRD peak was located at 3.152 Å$^{-1}$ and 3.144 Å$^{-1}$ for the 41 nm and 164 nm nanoparticles, respectively. This indicated that the 41 nm nanoparticles had a higher packing density than the 164 nm nanoparticles. Fig. 2b shows the ambient pressure PDFs that were generated using the ambient pressure diffraction data. The first peak for the PDF shows the nearest neighbor distance was 2.5 Å and 2.56 Å for 41 nm and 164 nm nanoparticles, respectively. The 41 nm nanoparticles had a smaller spacing between nearest neighbors which supported the result that the 41 nm nanoparticles have a higher packing density. The second peak and the third peak were at 4.01 Å and 4.81 Å, respectively, for the 41 nm nanoparticles and 4.12 Å and 4.93 Å, respectively, for the 164 nm nanoparticles. The average coordination number calculated from the PDF was 12.4 for 41 nm nanoparticles and 14.6 for 164 nm nanoparticles. This shows that the atomistic structure for the 41 nm nanoparticle is more densely packed than the 164 nm nanoparticles.

Fig. 3 shows the high pressure in situ XRD for 41 nm and 164 nm nanoparticles in neon quasi-hydrostatic pressure medium and the corresponding peak fitting. The XRD was collected while increasing pressure until ~55 GPa for 41 nm nanoparticles and until ~41 GPa for 164 nm nanoparticles. The maximum pressure was limited by the diamond culet size for each experiment. The amorphous pattern was preserved at all pressures indicating that the amorphous structure is very stable under high pressure compression for both nanoparticle sizes. The XRD peaks shifted towards the right with increasing pressure indicating a compression of the atomistic structure. Fig. 4a shows the position of the first XRD peak with increasing pressure for both 41 nm and 164 nm nanoparticles with corresponding lin ear line fits. T he first XRD peak shifted to higher q values by 7% and by 5% at the highest pressure for the 41 nm and 164 nm nanoparticles, respectively. At each pressure, the XRD peak for the 41 nm nanoparticle was at a higher value than the 164 nm nanoparticle. This indicated that the 41 nm nanoparticle was more compact than the 164 nm nanoparticle at each pressure. The position of the first XRD peak was used to calculate $[d_{max}(P)/d_{max}(0)]^3$ which is proportional to the reduced volume of the sample $V(P)/V(0)$. Here, P denotes high pressure and 0 denotes ambient pressure[25,26]. The $[d_{max}(P)/d_{max}(0)]^3$ decreased by 18% and 14% at the highest pressure for the 41 nm and 164 nm nanoparticles, respectively. The change in $[d_{max}(P)/d_{max}(0)]^3$ with pressure was used to fit a 3rd order Birch–Murnaghan equation of state[27] (see Fig. 4b-c). The calculated bulk modulus for the 41 nm nanoparticle was $208 \pm 28$ GPa and the pressure derivative of bulk modulus was $1.9 \pm 1$, and for the 164 nm nanoparticles the bulk modulus was $178 \pm 10$ GPa and the pressure derivative of bulk modulus was $6.5 \pm 1$. The slight increase in bulk modulus with decreasing nanoparticle size could be due to the increase in boron content.

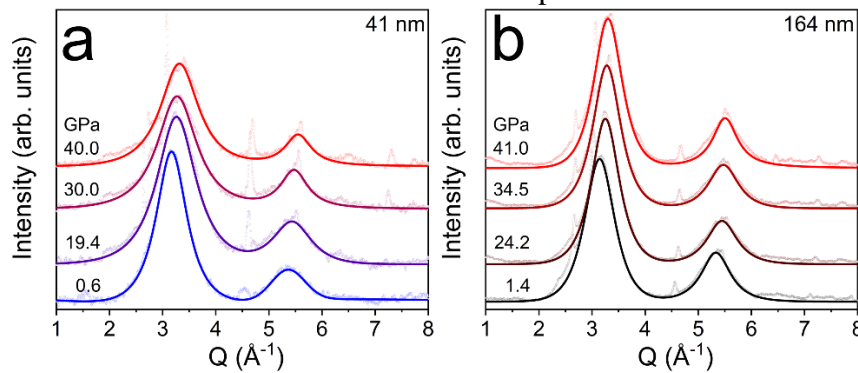

**Figure 3:** Experimental high pressure in situ XRD patterns and corresponding peak fitting for the amorphous peaks (with Re diffraction peaks removed) for a) 41 nm nanoparticles, and b) 164 nm nanoparticles.

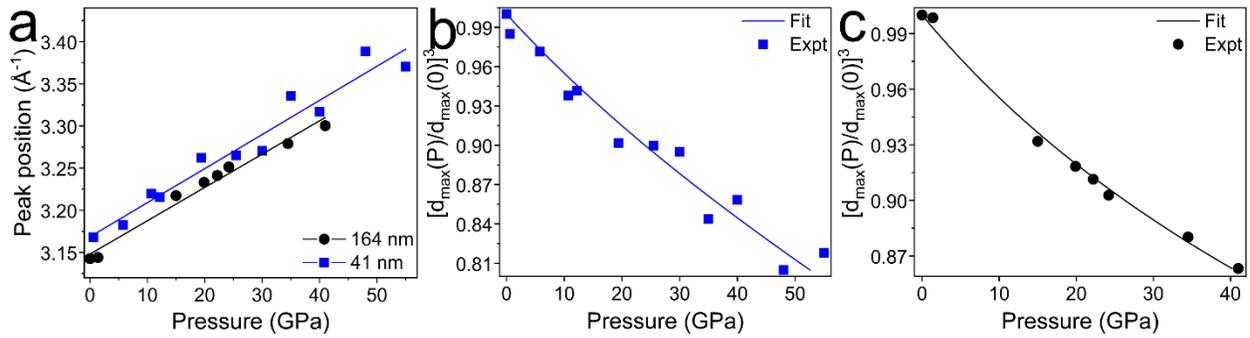

**Figure 4:** a) Peak position of the first amorphous peak with pressure for 41 nm and 164 nm sized $Ni_{1-x}B_x$ nanoparticles. $[d_{max}(P)/d_{max}(0)]^3$ volume change with pressure calculated from the first amorphous peak and the corresponding equation of state fit for b) 41 nm nanoparticles and c) 164 nm nanoparticles.

Fig. 5a-b shows the high-pressure PDF generated from the in situ XRD. There is a sudden drop in resolution for the high-pressure PDFs compared to ambient pressure PDFs (see Fig. 5a-b) due to the limited q-range accessible through a DAC. In addition, the signal to noise ratio increases with increasing pressure for high q-range which leads to a decrease in resolution with increasing pressure for the PDFs. This is most evident for the second and third peaks in the 41 nm nanoparticle PDF pattern, as the peaks merge with increasing pressure (Fig. 5a). The number of high-pressure PDF peaks and their shape remains similar with increasing pressure for the 164 nm nanoparticle. The PDF peaks shift towards the left with increasing pressure indicating a

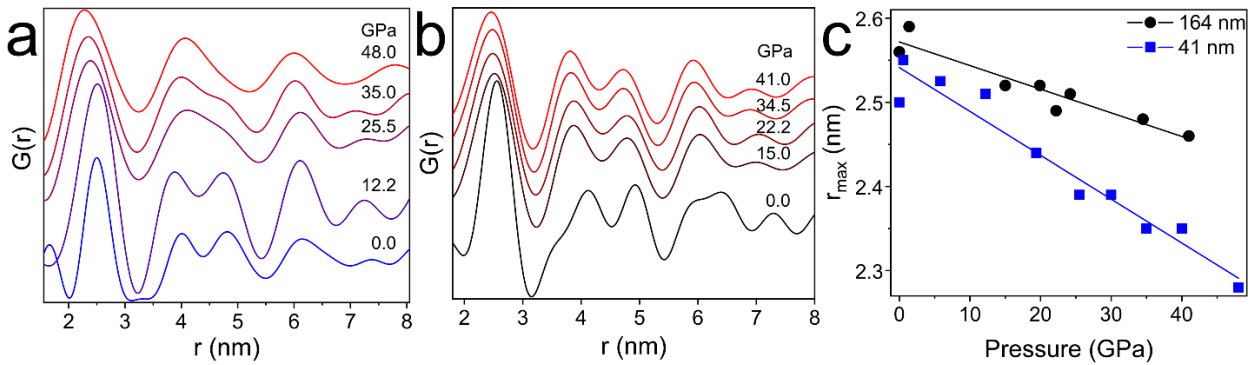

**Figure 5:** a) High pressure PDF for 41 nm $Ni_{1-x}B_x$ nanoparticles. b) High pressure PDF for 164 nm $Ni_{1-x}B_x$ nanoparticles. c) PDF first peak position with pressure for 41 nm and 164 nm $Ni_{1-x}B_x$ nanoparticles.

compression of the structure with pressure. The first PDF peak shifts by ~9% and ~4% at the maximum pressure for the 41 nm and 164 nm nanoparticles (see Fig. 5c). The first PDF peak for the 41 nm nanoparticles remains at a lower value than the 164 nm nanoparticles and also decreases more rapidly with pressure. The 41 nm nanoparticles had a lower pressure derivative of bulk modulus than the 164 nm nanoparticles which caused a more rapid decrease in the first peak position for the 41 nm nanoparticles with pressure.

## Discussion

Bulk $Ni_{0.81}B_{0.19}$ metallic glass PDFs have been generated using neutron diffraction and extended X-ray absorption fine structure methods[28–30]. These bulk results showed that the Ni-Ni, Ni-B and B-B nearest neighbor distances were 2.52 Å, 2.11 Å and 3.29 Å, respectively[28,29]. In contrast, XRD measurements used in this study are more sensitive to the Ni-Ni nearest neighbor distances than the Ni-B and the B-B nearest neighbor distances due to higher X-ray scattering from Ni atoms than B atoms. The observed nearest neighbor distance for the 164 nm nanoparticles with ~21% boron content was 2.56 Å, and for the 41 nm nanoparticles with a higher boron content of ~29% was 2.5 Å[28,29]. Kiani et al. used X-ray photoelectron spectroscopy to conclude that the nanoparticles were created through an aggregation of atomic clusters with B atoms at the center surrounded by a tightly packed Ni atom shell[14]. The remaining Ni atoms formed a loosely bound network connecting the different B centered atomic clusters. The Ni-Ni spacing for the Ni atoms in the B centered atomic cluster is smaller than the Ni-Ni spacing in the loosely bound Ni atom network connecting the B centered atomic clusters. The 41 nm nanoparticles had a higher boron content, which increased the density of B centered atomic clusters and reduced the observed average nearest neighbor distances for the Ni-Ni atoms. In contrast, the 164 nm nanoparticles had a larger number of Ni atoms in the loosely bound Ni atom network. This increased the observed average nearest neighbor distances for the Ni-Ni atoms. The maximum coordination number reported for the Ni-Ni nearest neighbors in bulk $Ni_{0.81}B_{0.19}$ was about 10.8. This is smaller than the observed coordination number of 12.4 and 14.6 for the 41 nm and 164 nm nanoparticles, respectively. This shows that the atomic arrangement within the $Ni_{1-x}B_x$ nanoparticles is different than the observed bulk structure. Reverse Monte Carlo simulations showed that the dominant Ni-B clustering in the bulk $Ni_{0.81}B_{0.19}$ was <0,3,6,0> Voronoi indices with a coordination number of 9, and the sample had Ni-B clusters ranging from 6 to 11 coordination numbers[28,31]. Iparraguirre et al. observed that the bulk $Ni_{0.81}B_{0.19}$ MG structure had distorted versions of clusters found in the $Ni_3B$ crystalline structure[32]. In contrast to conventional metallic glasses, the nanoparticles were synthesized through a bottom up approach via a growth process driven by the formation and aggregation of sub-nanometer sized nanoclusters. Our analysis shows that these nanoclusters can contain atomistic clusters with higher order symmetries, such as the icosahedral structure with 12 nearest neighbors. The nanocluster driven growth process could result in the higher coordination number observed for the $Ni_{1-x}B_x$ nanoparticles than for the bulk $Ni_{0.81}B_{0.19}$ MG.

The ratio of shear modulus (G) to bulk modulus (B) is an indicator of ductility in a MG[5]. A low ratio of G/B indicates a higher ductility. A low shear modulus and high bulk modulus is essential for improving ductility. Most bulk metallic glasses have low bulk modulus values ranging between 50-120 GPa[33] and only a few MGs have reported bulk modulus higher than 200 GPa like $Pd_{0.81}Si_{0.19}$ and $Pt_{0.6}Ni_{0.15}P_{0.25}$[4,33,34]. The bulk moduli measured for the 41 nm and 164 nm are larger than 170 GPa and match the trend calculated using the rule of mixtures. Using the rule of mixtures, the calculated bulk modulus is ~180 GPa and ~170 GPa for 41 nm and 164 nm nanoparticles, respectively (using ~160 GPa as the bulk modulus for crystalline Ni and ~224 GPa as the bulk modulus for crystalline boron[35]). The measured bulk modulus is lower than the bulk modulus for the crystalline $Ni_{1-x}B_x$ structures which ranges from 234-260 GPa[36]. The high pressure derivative of bulk modulus for the 164 nm nanoparticles could be due to the larger spacing between first nearest neighbors and a high coordination number at ambient pressure which allows for an easier rearrangement of atoms[37–40]. The high bulk modulus (>170 GPa), larger spacing between nearest neighbors for the 164 nm nanoparticles, and higher percentage of Ni-Ni metallic bonding due to lower boron concentration could result in the enhanced ductility of these large nanoparticles compared to smaller particles[14].

High pressure compression of $Ni_{1-x}B_x$ nanoparticles did not result in crystallization. The amorphous structure of the nanoparticles was stable up to the maximum pressure. An amorphous to amorphous phase transition (e.g. low density amorphous phase to high density amorphous phase[10,11]) did not occur. Researchers have reported only a few other stable amorphous structures at the same range of pressures. These include amorphous iron up to 67 GPa, amorphous $GeO_2$ up to 133 GPa, MG $Zr_{0.57}Cu_{0.154}Ni_{0.126}Al_{0.1}Nb_{0.05}$ up to 122 GPa, and $Pd_{0.4}Ni_{0.4}P_{0.2}$ up to 125 GPa [12,41–43]. The stability of these nanoscale MG structures show that they may be useful for use in extreme conditions.

**Conclusion**

In summary, 41 nm and 164 nm colloidally synthesized MG nanoparticles were compressed at high pressures. The ambient pressure PDFs showed that the 41 nm nanoparticle amorphous structure was more compact than the 164 nm nanoparticle. The nanoparticles were compressed using diamond anvil cell in a quasi-hydrostatic neon pressure medium. The bulk modulus was 208 GPa and 178 GPa for 41 nm and 164 nm nanoparticles, respectively. The difference in the ambient pressure structure and the bulk modulus for the two sizes was related to the difference in boron content. The amorphous structure was stable up to the maximum pressure of ~55 GPa for 41 nm nanoparticles and up to ~41 GPa for 164 nm nanoparticles, and showed no phase transition or crystallization due to high pressure. The high-pressure PDFs showed that the 41 nm nanoparticle structure became more compact with increasing pressure and the difference between the first PDF peak for the 41 nm and 164 nm nanoparticle structure increased with increasing pressure. These observations provide insight into the relationship between composition, atomic arrangement, deformation and mechanical properties in metallic glasses. These ultrastable MG nanoparticles could be incorporated into a metal or ceramic matrix composite as strengtheners and could contribute additional functionality such as catalytic activity[44–47].

**Acknowledgement**

We thank Prof. Wendy Mao and Dr. Feng Ke at Stanford University for sharing their resources for drilling the sample chamber. We thank David Doan, Andrew Doran and Dr. Martin Kunz for assistance with initial experiments. We also acknowledge the assistance of Sergey N. Tkachev with neon gas loading for quasi-hydrostatic diamond anvil cell measurements. Part of this work was performed at GeoSoilEnviroCARS (The University of Chicago, Sector 13), Advanced Photon Source (APS), Argonne National Laboratory. GeoSoilEnviroCARS is supported by the National Science Foundation – Earth Sciences (EAR – 1634415) and Department of Energy- GeoSciences (DE-FG02-94ER14466). This research used resources of the Advanced Photon Source, a U.S. Department of Energy (DOE) Office of Science User Facility operated for the DOE Office of Science by Argonne National Laboratory under Contract No. DE-AC02-06CH11357. Use of the COMPRES-GSECARS gas loading system was supported by COMPRES under NSF Cooperative Agreement EAR -1606856 and by GSECARS through NSF grant EAR-1634415 and DOE grant DE-FG02-94ER14466. Portions of this work were performed at HPCAT (Sector 16), Advanced Photon Source (APS), Argonne National Laboratory. HPCAT operations are supported by DOE-NNSA's Office of Experimental Sciences. Part of this work was performed at the Stanford Nano Shared Facilities (SNSF), supported by the National Science

Foundation under award ECCS-1542152. A.C.L is supported by Knight-Hennessy Scholars. M.T.K. is supported by the National Defense and Science Engineering Graduate Fellowship. We gratefully acknowledge financial support from the grant DE-SC0021075 funded by the U.S. Department of Energy, Office of Science